\documentclass[namedreferences]{SolarPhysics}
\usepackage[optionalrh]{spr-sola-addons} % For Solar Physics 
\usepackage{graphicx}        % For eps figures, newer & more powerfull
\usepackage{color}           % For color text: \color command
\usepackage{url}             % For breaking URLs easily trough lines
\newcommand{\aap}{    {\it Astron. Astrophys.}}
\newcommand{\aaps}{   {\it Astron. Astrophys. Suppl.}}

\newcommand{\apj}{    {\it Astrophys. J.}}
\newcommand{\apjs}{   {\it Astrophys. J. Suppl.}}

\newcommand{\grl}{    {\it Geophys. Res. Lett.}}

\newcommand{\solphys}{{\it Solar Phys.}}

\begin{document}

\begin{article}

\begin{opening}

\title{Solar EUV Spectrum Calculated for Quiet Sun Conditions}
     
\author{M.~\surname{Haberreiter}}
\runningauthor{Haberreiter}
\runningtitle{Calculation of the Solar EUV Spectrum}

   \institute{$^{1}$ Physikalisch-Meteorologisches Observatorium Davos, World Radiation Center, Dorfstrasse 33, 7260 Davos Dorf, Switzerland
                     email: \url{margit.haberreiter@pmodwrc.ch} \\ 
              $^{2}$ LASP, University of Colorado, 1234 Innovation Drive, Boulder, CO, 80303, USA
             }

\begin{abstract}
We present spectral synthesis calculations of the solar extreme UV (EUV) in spherical symmetry carried out with the 'Solar Modeling in 3D' code. The calculations are based on one-dimensional atmospheric structures that represent a temporal and spatial mean of the chromosphere, transition region, and corona. The synthetic irradiance spectra are compared with the recent calibration spectrum taken with the EUV Variability Experiment during the Whole Heliospheric Interval. The good agreement between the synthetic and observed quiet Sun spectrum shows that the employed atmospheric structures are suitable for irradiance calculations. The validation of the quiet Sun spectrum for the present solar minimum is the first step towards the modeling of the EUV variations. 
\end{abstract}


\end{opening}

\section{Introduction}
The solar EUV irradiance changes on short time-scales of minutes to hours, as well as longer times-scales such as the 27-day solar rotation period and the 11-year solar cycle. Since the peculiarly low 2008 solar minimum there are also indications that the EUV shows a long-term trend. For example, the SOHO/SEM measurements which cover a full solar cycle show a decrease of 15\,\% between the 1996 and 2008 solar minimum \cite{Didkovsky2010ASPC}. 

Incident EUV radiation is absorbed by the Earth's thermosphere, which leads to partial ionization - forming the ionosphere - and a change of its temperature and density. As shown by \inlinecite{Solomon2010} the EUV is in fact the main driver of density changes in the ionosphere. It is understood that the low EUV irradiance during the 2008 minimum phase led to an unprecedented decrease of the ionospheric density \cite{Emmert2010GeoRL}. Furthermore, as the density of the upper atmosphere has a direct effect on satellite drag \cite{FullerRowell2004,Lilensten2008AnGeo} it is clear that models that predict the thermospheric density require an improved knowledge of the incident EUV radiation.

Modeling the EUV coronal lines have already been carried out using different approaches. The most well-known models are the CHIANTI models of differential emission measures \cite[DEMs]{Chianti1,Chianti2} which allow calculation of the optically thin emissions of the quiet and active corona, and coronal holes. \inlinecite{Sirk2010} follow this approach, but use a set of isothermal plasmas in order to reproduce observations of the quiet Sun. However, the approach using DEMs is limited to modeling the pure optically thin emissions from the corona. As already pointed out by \inlinecite{Judge1995ApJL} and also discussed in Section\,\ref{sec:spherical} this approach is not always sufficient, as it depends on the opacity of the spectral line. 

Other reconstruction approaches involve the use of proxies to describe the EUV variability. A well-known example is SOLAR2000 \cite{Tobiska2000}. There is still ongoing work to determine which proxies, or spectral lines best represent the variations of the entire EUV spectrum \cite{Kretzschmar2008AcGeo,deWit2009}. Proxy models have been quite successful in describing the EUV variations, however, there are indications that they fail to reproduce the low EUV values of the 2008 minimum. Finally, the NRLEUV model \cite{NRLEUV} is based on EUV images and calculates the emergent spectrum from the CHIANTI models. 

Our approach to model EUV irradiance variations is to calculate intensity spectra for different features on the solar disk, as identified from various images of the solar disk. Depending on the wavelength range the EUV shows contributions from the chromosphere, transition region (TR), and corona. Therefore, both the optically thick emissions from chromospheric and TR layers, as well as the mainly optically thin emissions from the corona need to be included. 

In the next section we give details about the spectral synthesis code Solar Modeling in 3D (SolMod3D), the employed semi-em\-pi\-rical atmospheric structures, in particular for the corona, and the spherical integration scheme. Then, in Section\,\ref{sec:results} we compare the calculations for the quiet Sun with the spectrum measured with the EUV Variability Experiment \cite[EVE]{Woods2006} during a rocket calibration flight \cite{Chamberlin2009}. Finally, we discuss the results and give an outlook for the reconstruction of the spectral irradiance for the full solar cycle.

\section{Reconstruction approach}
Our approach to model spectral irradiance variations is to calculate intensity spectra for different features on the solar disk. The features are identified from intensity images taken with the Precision Photometric Solar Telescope \cite[PSPT]{PSPT} and the Extreme UV Imaging Telescope \cite[EIT]{EIT1995} onboard the ESA/NASA satellite SOHO. Weighting the intensity spectra with the time-dependent relative area covered by the features yields the time-dependent spectral irradiance. For visible and UV wavelength ranges similar approaches have already been employed by \inlinecite{FontenlaHarder2005}, \inlinecite{Harder2005}, and \inlinecite{Haberreiter2005ASpR}. Here, we extend this approach to the EUV. 

\subsection{The SolMod3D code} \label{sec:code}
The spectral synthesis of the EUV is carried out with the SolMod3D code. It is a branch of the Solar Radiation Physical Modeling project \cite{Fontenla2009ApJ,Fontenla2007,Fontenla2006}, a state-of-the-art radiative transfer code in full non-local thermodynamic equilibrium (NLTE). SolMod3D has been recently updated to spherical line-of-sight integration \cite{HaberreiterFontenla2009AIPC} and applied in solar limb studies \cite{Thuillier2010}. In comparison to the plane-parallel geometry, the spherical symmetry leads to an increase of the irradiance in some lines by up to a factor of two. This increase is due to the inclusion of the extended corona, which cannot be accounted for with a plane-parallel approach. Thus, the spherical symmetry is essential for the correct calculation of the coronal emission.

\subsection{Atmospheric structures}
Our purpose is to calculate the EUV from 10 to 100\,nm. Depending on the wavelength the formation of the EUV ranges from the chromosphere, TR, and corona. Therefore, the calculations are based on semi-em\-pi\-rical structures that represent these layers in the solar atmosphere.

For the photosphere, chromosphere, and TR we employ the latest semi-em\-pi\-rical atmospheric structures by \inlinecite{Fontenla2009ApJ} and solve the full NLTE radiative transfer for the most abundant elements from hydrogen to iron up to an ion charge of two. For the coronal structure the temperature gradient has been derived by semi-em\-pi\-ri\-cally adjusting temperature values from eclipse observations by \inlinecite{Singh1982}. The density values are extended from the TR to the corona based on the values given by \inlinecite{Doschek1997}. Figure~1 gives our first version of these coronal models that allow the observed quiet-Sun EUV spectrum to be reproduced very well. The main variables of the model atmospheric structures such as temperature, electron density, proton density, and pressure can be obtained from http://www.pmodwrc.ch/pmod.php?topic=sci. Currently, the structures are purely semi-empirical, so hydrostatic equilibrium has not yet been validated. In a future version we will consider the radiative losses and also see whether the hydrostatic equilibrium can be applied to the models. Details on the models will continuously be updated on the above-mentioned webpage.

For the coronal structures the statistical equilibrium equation is solved for ionization stages three and higher with the optically thin approach. The atomic data is taken from the NIST database, the Opacity Project, and the Chianti database, Version 5.4 \cite{Chianti1}. At present, we consider a total of approx. 14,000 atomic levels and 170,000 spectral lines. We are aware that these models cannot reproduce the rich dynamics of the corona. In fact, these semi-em\-pi\-rical structures should be considered to be a temporal and spatial mean of the extended corona suited for the purpose of reconstructing solar irradiance.   

\begin{figure}[tt]
\vspace{1cm}
\centering
\includegraphics[width=.55\linewidth]{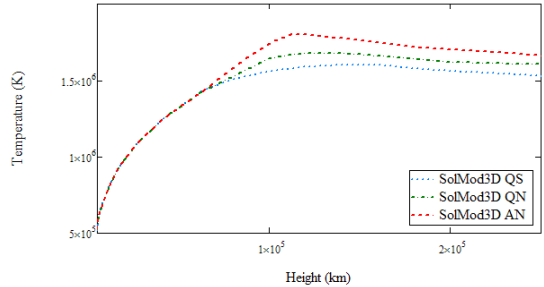}
\caption{Coronal models for the quiet Sun inter-network (QS), the quiet network (QN), and the active network (AN) used in SolMod3D for the calculation of the coronal emission lines.  \label{fig:models}} 
\end{figure}

\subsection{Spherical Symmetry}\label{sec:spherical}
The spherical geometry in SolMod3D allows realistic line-of-sight intensities at and beyond the solar limb to be calculated. This geometry has already been used in the COde for Solar Irradiance \cite[COSI]{Haberreiter2008b} to conduct solar limb studies \cite{Haberreiter2008a,Thuillier2010}. The scheme is essential for the calculation of the coronal spectrum, as a considerable part of the irradiance is emitted by the extended corona. Even more important is the fact that due to the increased line-of-sight at and beyond the solar limb, opacity effects are not always negligible. In fact we find that for some spectral lines the maximal optical depth can reach values of the order of 0.3 to 0.9. 

\begin{figure}[t!]
\centering
\includegraphics[width=.76\linewidth]{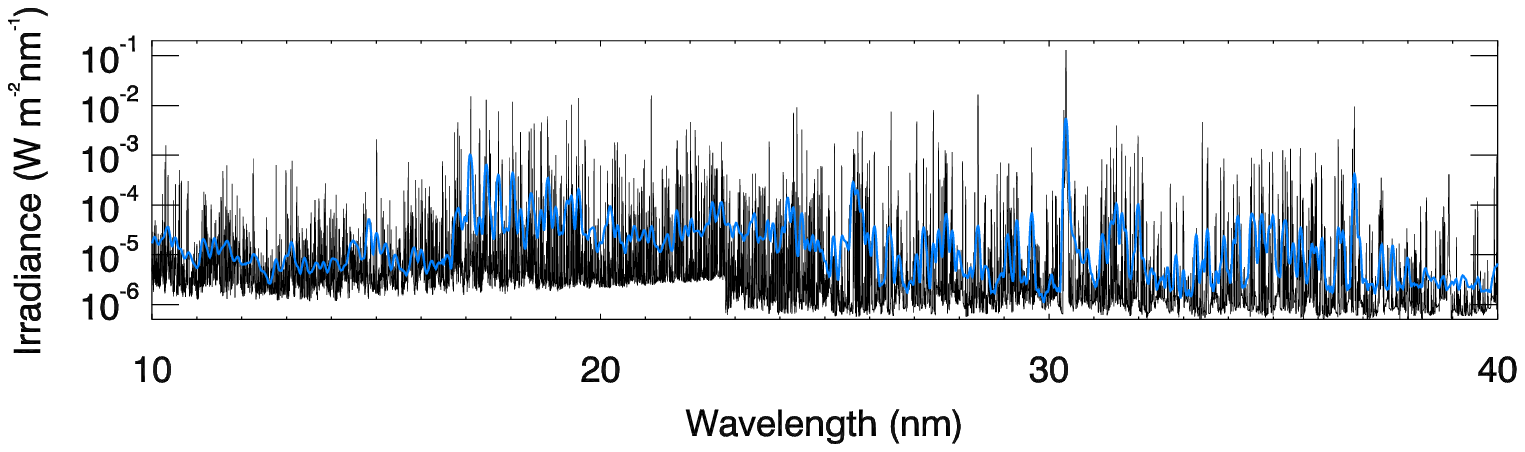}
\includegraphics[width=.76\linewidth]{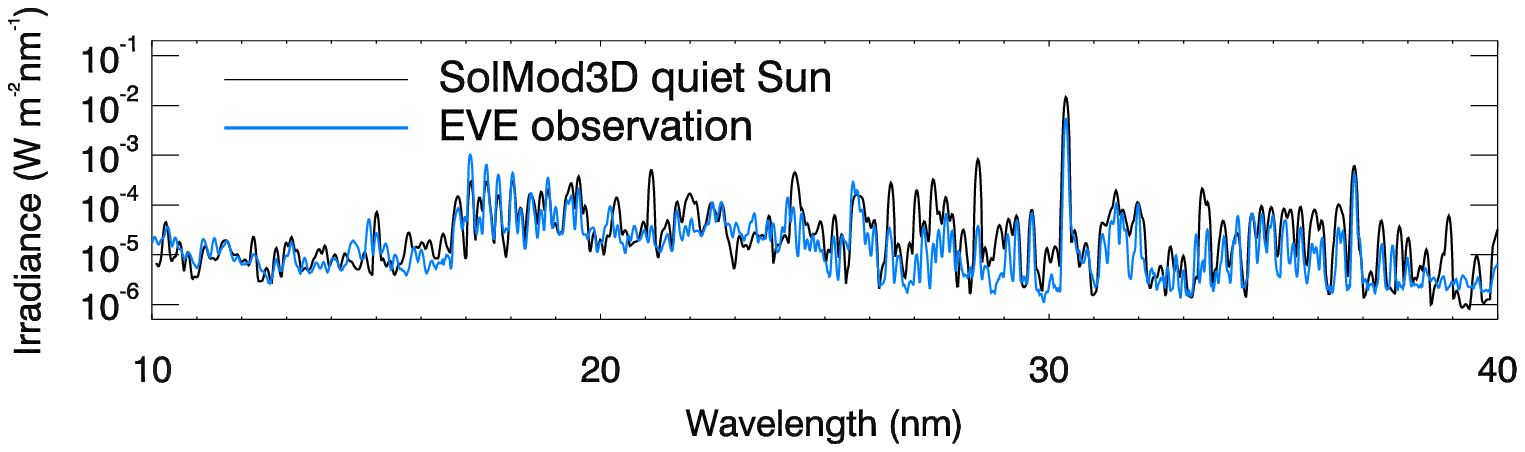}
\includegraphics[width=.76\linewidth]{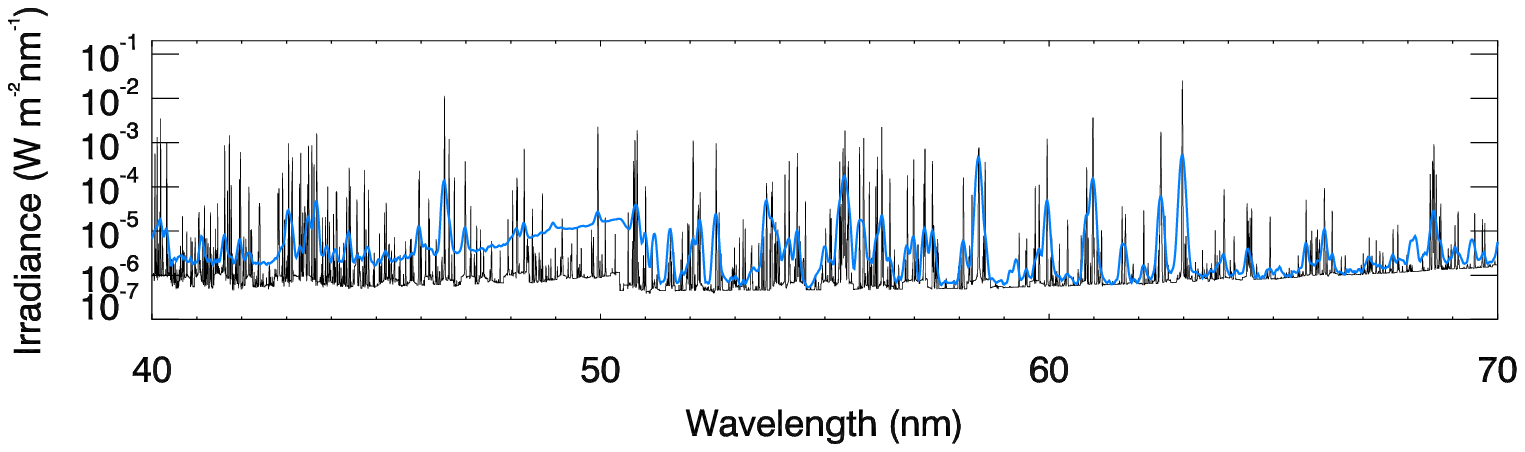}
\includegraphics[width=.76\linewidth]{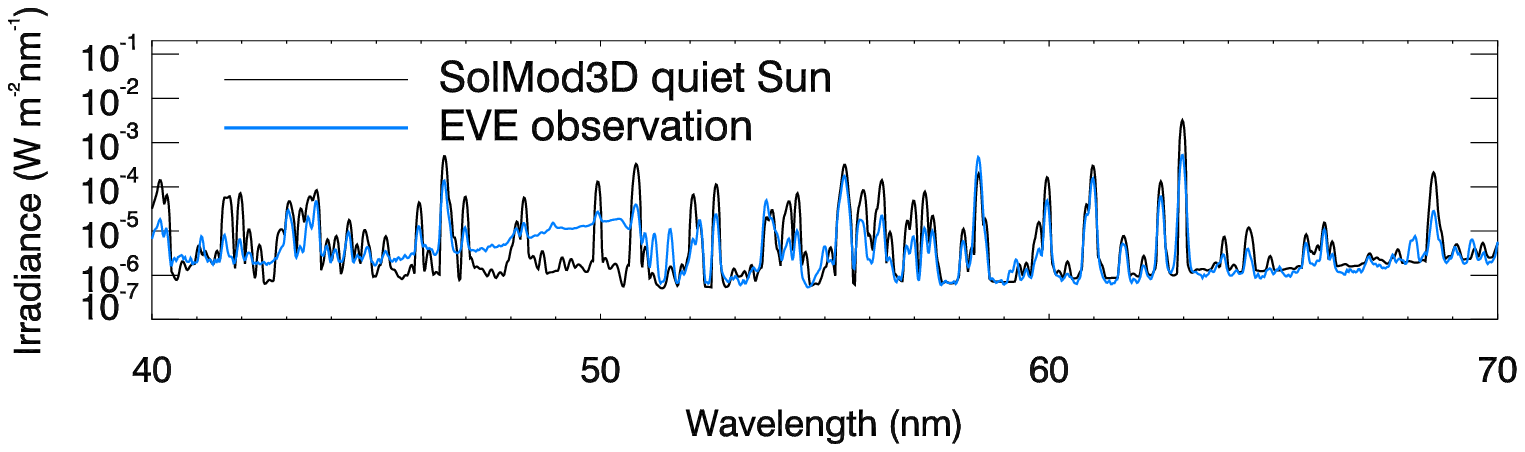}
\includegraphics[width=.76\linewidth]{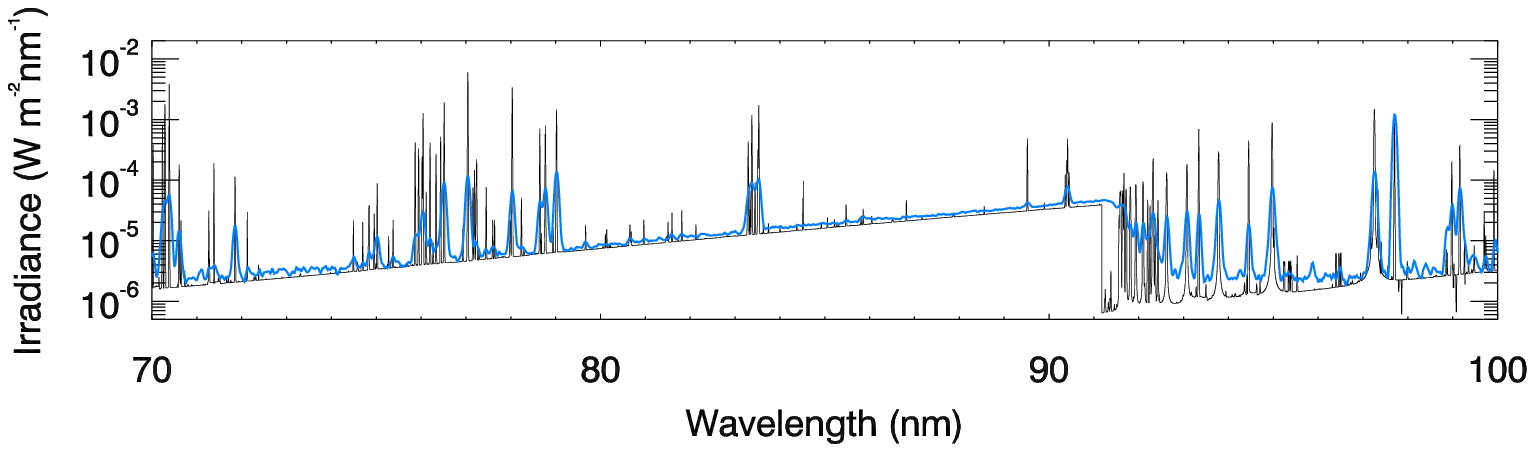}
\includegraphics[width=.76\linewidth]{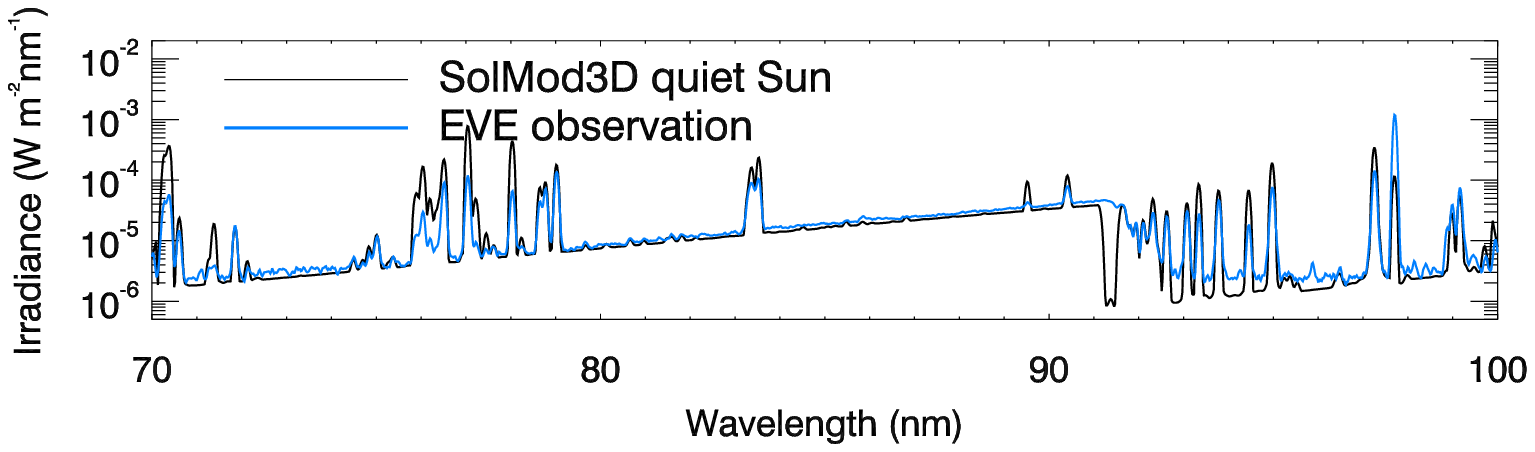}
\caption{Comparison of the synthetic spectrum calculated with SolMod3D. The first, third, and fifth panels show the high-resolution synthetic spectrum (black line) compared with the spectrum observed with the EVE instrument during a rocket flight on April 14, 2008 (blue thick line). The second, fourth, and sixth panels show the synthetic spectrum (black line) convolved twice with a 1-{\AA} boxcar compared with the EVE rocket spectrum (blue thick line).  \label{fig:eve}} 
\end{figure}

\section{Results}\label{sec:results}
The first, third, and fifth panels of Figure~2 show the calculated EUV spectrum in high resolution for the quiet Sun compared with the spectrum taken with the EVE calibration instrument during a rocket flight on April 14, 2008 \cite{Chamberlin2009}. This EVE calibration campaign was set within the time frame of Whole Heliospheric Interval (WHI). As the Sun was almost free of any active regions on the day of the observation, the spectrum practically represents the quiet Sun. The synthetic quiet Sun spectrum is derived using a combination of {75\,\%} of the spectrum calculated for the quiet Sun inter-network, {22\,\%} of quiet network, and {3\,\%} of active network. This notation follows the one used by \inlinecite{Fontenla2009ApJ}. This combination is a typical distribution of features for quiet Sun conditions. Currently, we are in the process of a more detailed analysis of EIT and the corresponding PSPT images to quantify the contributions of various activity features over the full solar cycle. 

The high-resolution spectra show much more detail compared to the observed spectrum. This might be important for understanding the variability in the EUV. In particular, it is to be expected that not all spectral lines vary to the same degree over the solar cycle. Therefore, it is essential to calculate the solar spectrum in full detail. This comparison also shows that it is of great importance to include the complete atomic data for the calculation of the EUV spectrum. 

The convolved synthetic spectra, given in the second, fourth, and sixth panels of Figure~2, show satisfactory agreement with the observations. This indicates that the coronal models, applied to represent solar minimum conditions, are reasonable representations of the physical parameters of the corona. Nevertheless, in a few spectral ranges the SolMod3D calculation leads to discrepancies from the observations. Most prominent are the underestimation of the He\,{\sc ii} edge at 50.4\,nm. One reason for this could be that helium diffusion is not included in the models. Also, the temperature gradient of the transition region might need to be re-evaluated. Another slight discrepancy can also be seen in the pseudo-continuum just long ward of the H\,{\sc i} edge at 91.1\,nm, which is not well reproduced in the calculation due to the limited number of hydrogen levels.

\section{Conclusions}
Solar spectral irradiance variations in the EUV are important for the detailed modeling of the Earth's ionosphere. Based on 1D atmospheric structures we are able to reproduce the quiet Sun EUV irradiance spectrum. This shows that the employed atmospheric structures are a realistic representation of the coronal plasma. In a next step we will calculate spectra for the active components of the corona and coronal holes. Finally, based on the analysis of solar images, following {\em e.g.} \inlinecite{Barra2009}, we intend to model the variations over the full EUV range. This will hopefully give us some insight into why the EUV irradiance was considerably lower during the 2008 minimum than the previous minimum. This work is also important for the interpretation of the detailed observations by SDO/EVE and also PROAB2/LYRA \cite{LYRA2006}.
%--------------------------------------------
{\begin{acks}
We are grateful to Juan Fontenla for many fruitful discussions on radiative transfer and for providing the first version of coronal atmospheric structures. We thank Phil Chamberlin for providing the EVE calibration spectrum. We also thank Tom Woods and the Solar Influences group at LASP for detailed discussions, as well as Eugene Avrett for his valuable comments as reviewer of the manuscript. MH acknowledges support by the Swiss Holcim Foundation for the Advancement of Scientific Research. 
\end{acks}}

\end{article} 
\end{document}